\documentclass[nato,noid,numreferences]{crckbked}

\usepackage{graphicx}
\usepackage{epsf}
\newcommand{\be}{\begin{equation}}
\newcommand{\en}{\end{equation}}
\newcommand{\bea}{\begin{eqnarray}}
\newcommand{\ena}{\end{eqnarray}}
\newcommand{\hbo}{\hbox to 1 true cm {\hfill } }

\begin{document}

\begin{opening}
\title{VORTEX INDUCED CONFINEMENT AND THE \protect\\
       KUGO-OJIMA CONFINEMENT CRITERION}



\author{Kurt Langfeld}
\institute{Insitut f\"ur Theortische Physik, Universit\"at T\"ubingen \\
           Auf der Morgenstelle 14, D-72076 T\"ubingen }



\begin{abstract}
The SU(2) Yang-Mills gluon and ghost Green-functions are studied in Landau 
gauge by means of lattice gauge simulations. A focal point is their 
low energy behavior since in particular the Kugo-Ojima confinement criterion 
relates a diverging ghost form factor at vanishing momentum transfer 
to the confining capabilities of the theory. This divergence is verified by 
numerical simulations. Removing the confining vortices from the lattice 
ensembles converts SU(2) Yang-Mills theory in a non-confining theory. 
It is shown that in this modified theory the divergence of the ghost 
propagator disappears. 
\end{abstract}

\end{opening}

\section{Introduction}

The understanding of confinement of the fundamental degrees of
freedom, the quarks and the gluons, is still one of the big challenges of QCD. 
In the case of heavy quarks, a signal of quark confinement can be 
deduced from the static quark anti-quark potential~\cite{bali95}, which 
is linearly rising at low temperatures. There is only an indirect 
evidence for the confinement of gluons: there is no gluonic contribution 
to the vacuum energy density for temperatures below the critical one, 
while above the critical temperature the gluonic blackbody 
radiation dominates~\cite{eng95}. In the case of dynamical quarks with 
a small but non-vanishing quark current mass a precise definition of 
quark confinement is cumbersome. For these reason, a litmus paper 
which signals the absence of colored states from physical correlation 
functions would be highly desirable. 

\vskip 0.3cm 
In the past, it has been argued that the low energy behavior of 
QCD Green-functions in Landau gauge encodes the information on 
confinement~\cite{zwa92,kugo79}: Zwanziger argued that 
in the confinement phase the gauge field configurations which are relevant 
in the thermodynamic limit are restricted to the Gribov horizon~\cite{zwa92}. 
In this case, a ghost propagator which diverges at zero momentum 
transfer reports that the above horizon condition is satisfied. In the 
framework of the BRST quantization, Kugo and Ojima proposed a criterion 
which signals that the physical subspace only consists of color singlet 
states~\cite{kugo79}. The general version of this criterion involves 
a certain Green-function which approaches $-1$ in the zero momentum 
limit. Recent lattice investigations which challange this general
version can be found in~\cite{nak01}. It is easier from a technical 
point of view to test the criterion in Landau gauge. 
In this gauge, this criterion is fulfilled if the 
ghost form factor is singular at zero momentum~\cite{kugo95}. It therefore 
coincides with Zwanziger's horizon condition. In the derivation of the 
Kugo-Ojima criterion, one has assumed that a BRST charge operator 
is uniquely defined for the whole configuration space. At the present 
stage of investigations, this assumption is unjustified due to the 
presence of Gribov ambiguities~\footnote{ I thank Pierre Van Baal and 
Maarten Golterman for helpful discussions.}. 

\vskip 0.3cm 
In order to explore the relation between the infrared behavior of 
Yang-Mills Green-functions and confinement, the vortex picture of 
confinement is a convenient tool at our disposal. Removing 
the confining vortices~\cite{deb97} (see below for details) from 
the lattice configurations turns the Yang-Mills theory into 
a (presumably non-local) gauge theory which does not confine quarks. 

\vskip 0.3cm 
In this paper, I will compare the gluon and ghost form factors of the 
pure SU(2) gauge theory with those of the modified (non-confining) theory. 
Firstly, this will provide information on the signature of 
confinement in these correlations functions. Secondly, since these 
Green-functions are important ingredients for hadron 
phenomenology~\cite{alk01}, a knowledge of the vortex impact on these 
Green-functions will help to understand the role of the confining 
vortices in hadron physics. 

\section{ Approaching low energy Yang-Mills theory } 
\subsection{ The vortex picture } 

The basic idea to get insights into the mechanism of quark confinement 
is to project the SU(2) lattice configurations onto fields belonging 
to a simpler gauge group such as $Z_2$. The ($Z_2$--gauge invariant) 
degrees of freedom of the latter theory are closed surfaces in 
four space time dimensions. Equivalently, at a given time slice 
these surfaces can be viewed as closed strings in the spatial hypercube, 
i.e.~the so-called center vortices. The difficulty 
is to find the particular definition of these objects which ensures that 
they are sensible degrees of freedom in the continuum limit. A progress 
in this direction was made in the recent past~\cite{deb97}: a new gauge, 
the so-called Maximal Center Gauge (MCG) was invented. After MCG gauge 
fixing, the vortex degrees of freedom are defined by projecting full 
configurations onto vortex ones. Subsequently, a tight relation of these 
vortices to the physics of confinement was established: the MCG vortex 
ensembles reproduce the confining quark potential, while a 
removal (see e.g.~\cite{la02}) 
of these vortices from the lattice ensembles results in a Coulomb type 
potential. Using this particular gauge, it turned out that the 
vortices are indeed sensible degrees of freedom in the continuum 
limit~\cite{la98}. In addition, an intriguing picture of the deconfinement 
phase transition is available, which appears as a vortex de-percolation 
transition~\cite{la99}. 

\vskip 0.3cm 
Here, I will use the lattice ensembles from which the MCG vortices have 
been removed as a model theory which does not show quark confinement. 
This theory which is defined by the modified lattice ensembles 
possesses the following properties: (i) it is SU(2) gauge invariant; 
(ii) it is presumably non-local, since non-local field configurations 
have been removed from a local theory; 
(iii) since the objects which have been removed 
exist on a physical length scale, the renormalization group flow 
towards the continuum limit is the same as the one of the full theory. 

\subsection{ Gluon and ghost form factors } 

In the following, the renormlazied gluon and ghost propagators,  
$D^{ab}_{\mu \nu }(p)$ and $G^{ab}(p)$, are parameterized in Landau 
gauge by 
\be 
D^{ab}_{\mu \nu }(p) \; = \; \delta ^{ab} P_{\mu \nu } \; 
\frac{F(p^2,\mu^2)}{p^2} \; , 
\hbo 
G^{ab}(p) \; = \; \delta ^{ab} \; \frac{ G(p^2,\mu^2 ) }{p^2} \; , 
\label{eq:1} 
\en 
where $p$ is the momentum transfer, $P_{\mu \nu }$ is the transverse 
projector and $\mu $ the remormalization point where
$F(\mu^2,\mu^2)=1$, $G(\mu^2,\mu^2)=1$. 
By means of a truncated set of Dyson-Schwinger equations, 
the form factors in (\ref{eq:1}) have only 
recently been addressed~\cite{sme97,blo01}. It was pointed out that, 
at least for a certain truncation scheme, the gluon and ghost form factors 
satisfy simple scaling laws in the infra-red momentum range, 
in particular (for $p^2 \ll  (1 \, \mathrm{GeV})^2 $)
\be 
F(p^2,\mu^2) \; \propto \left[ p^2 / \mu^2 \right]^{2 \kappa } \; , \hbo 
G(p^2,\mu^2) \; \propto \left[ p^2 / \mu^2 \right]^{- \kappa } \; . 
\label{eq:2}  
\en 
Depending on the truncation of the Dyson tower of equations, one finds 
$\kappa $ ranging from $0.5$ to $0.7$~\cite{ler02,fi02}. These findings 
match with Zwanziger's horizon condition: the ghost form factor is singular 
in the limit $p \rightarrow 0$. 

Of great importance for phenomenological purposes is the  running coupling 
strength $\alpha (p^2) $. This strength directly enters the quark 
DSE~\cite{blo02}, and can be interpreted as an effective interaction strength 
between quarks. Surprisingly, the prediction that the running coupling 
strength in Landau gauge~\cite{sme97,blo01}
\be 
\alpha (p^2) \; = \; \frac{ \widetilde{Z}^2_1 (\mu ^2, \Lambda ) }{ 
\widetilde{Z}^2_1 (p^2 , \Lambda ) } \; 
\alpha (\mu^2) \, F(p^2,\mu ^2) \, G^2(p^2, \mu^2) \; , 
\label{eq:12} 
\en 
approaches a constant in the limit $p^2 \rightarrow 0$ is independent 
of the truncation and approximations used in~\cite{sme97,blo01}. 
In the DSE approach the ghost-gluon vertex renormalization constant 
$\widetilde{Z}_1$ is set to unity. This is assumed since 
$\widetilde{Z}_1$ is finite at least to all orders of perturbation 
theory~\cite{tay71}. 

\begin{figure}[tbh] 
\centerline{ 
\epsfxsize=6cm\epsffile{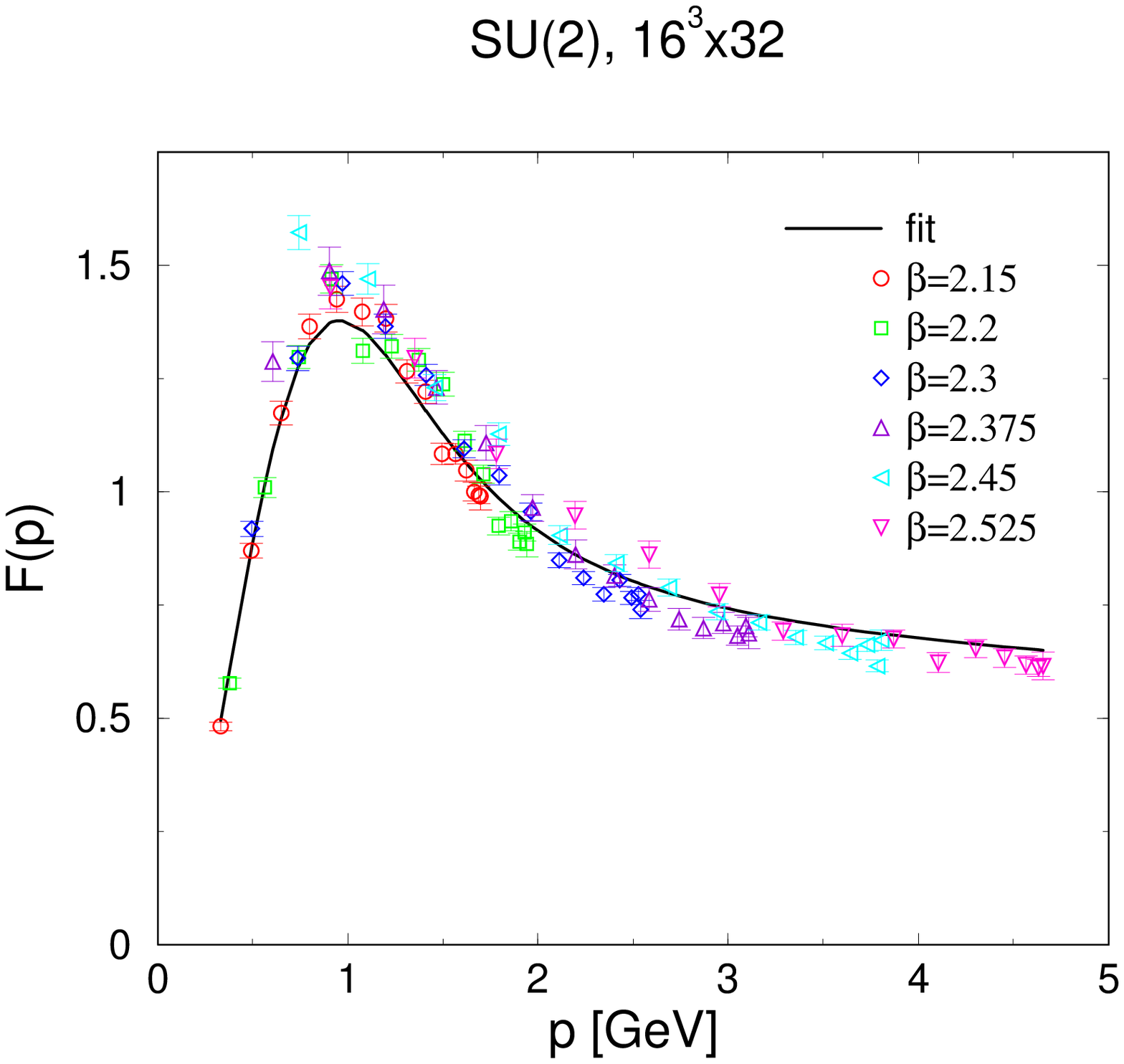}
\epsfxsize=5.7cm\epsffile{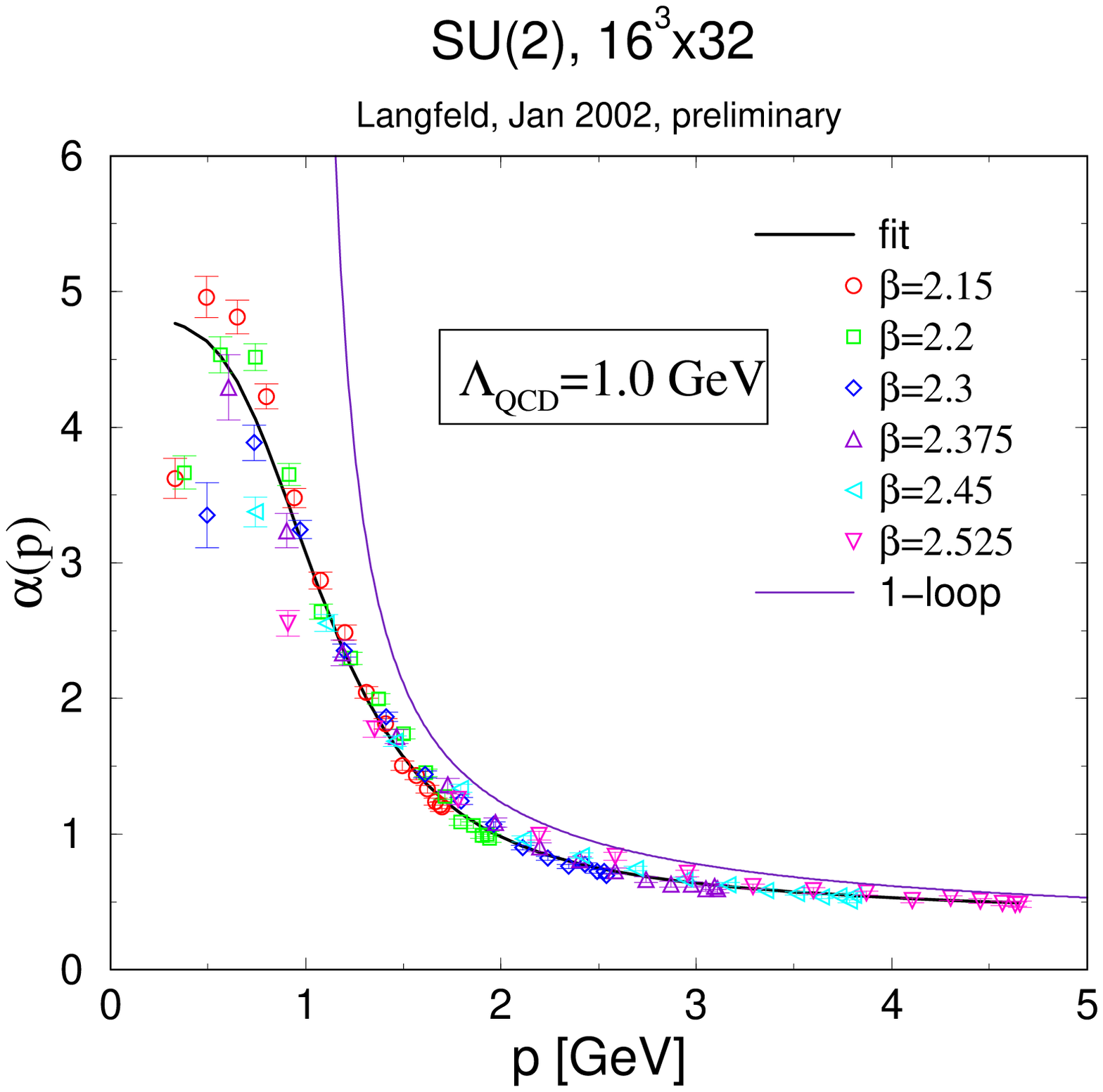}
}
\caption{ The gluon form factor (left panel) and the running coupling 
   strength (right panel) as function of the momentum transfer; pure SU(2) 
   gauge theory.
}\label{fig:1}
\end{figure}
\section{ Numerical results } 

\subsection{ Pure SU(2) theory } 

The simulations were carried out on a $16^3 \times 32 $ lattice 
for $\beta $ values ranging from $2.1$ to $2.5$. Physical units 
are obtained by eliminating the lattice spacing $a$ using the scaling 
relation 
\be 
\sigma \, a^2(\beta ) \; = \; 0.12 \, \exp \left\{ - 
\frac{ 6 \pi ^2 }{11} \left( \beta - 2.3 \right) \right\} \; , 
\label{eq:3} 
\en 
where I used the string tension $\sigma = (440 \mathrm{MeV} )^2 $ as 
a reference scale. The form factors were 
directly calculated from appropriate correlation functions 
(rather than calculating the propagator times $p^2$). This method 
efficiently suppresses the statistical noise~\cite{la02}. 
For a proper definition of the gluon fields see~\cite{la02},  and for the 
definition of the ghost propagator see~\cite{su96}. The Landau gauge 
condition  was implemented using a simulated 
annealing algorithm~\cite{su96}. 

Initially, I obtain the un-renormalized form factors 
$F_B(p^2, \beta )$, $G_B(p^2, \beta )$ as function of the momentum 
in physical units. Note that in view of (\ref{eq:3}) $\beta \propto 
\ln ( \Lambda /\sqrt{\sigma }) $ encodes the logarithmic dependence on 
the cutoff $\Lambda = \pi / a$. The desired 
{\tt renormalized } form factors are obtained via multiplicative 
renormalization, i.e. 
\be 
F(p^2,\mu^2) 
\; = \; Z_3^{-1}(\mu ^2, \beta ) \;  F_B(p^2, \beta ) \; , \; \; 
G(p^2, \mu ^2 ) 
\; = \; \widetilde{Z}_3^{-1}(\mu ^2, \beta ) \;  G_B(p^2, \beta ) \; . 
\label{eq:4} 
\en 
In the figures, the overall normalizations of the gluon and the ghost form
factor, respectively, were arbitrarily chosen. 

\begin{figure}[tbh] 
\centerline{ 
\epsfxsize=6.5cm\epsffile{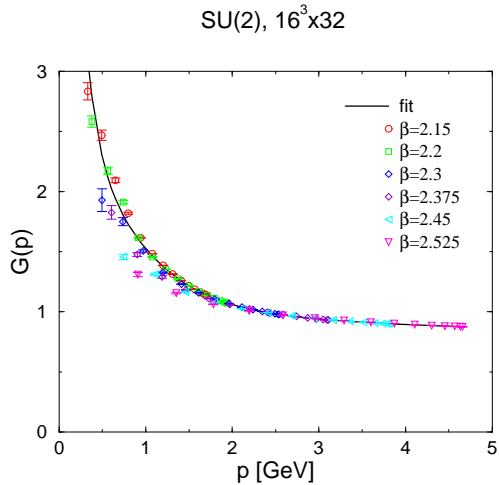}
}
\caption{ The ghost form factor 
   as function of the momentum transfer; pure SU(2) gauge theory.
}\label{fig:2}
\end{figure}
The ghost-gluon vertex renormalization constant $\widetilde{Z}_1$ can
be obtained by demanding 
\be 
\widetilde{Z}_1^{-2} (\mu ^2, \beta )  \; \alpha _0 \, F_B(p^2, \beta) 
\, G_B^2(p^2, \beta ) \rightarrow \mathrm{finite} \; , \hbo \forall 
\; p \; ,
\label{eq:5} 
\en 
where $\alpha _0 \propto 1/\beta $ is the bare gauge coupling constant 
squared. 
In view of~\cite{tay71}, one expects that $\widetilde{Z}_1(\mu ^2, \beta ) 
\rightarrow $ constant in the lattice calculations  
when the continuum limit is reached 
$\beta \rightarrow \infty $. In practical lattice calculations using 
the Wilson action, it turned out that a significant dependence 
of $\widetilde{Z}_1$ on $\beta $ is present for $\beta < 2.3$. 
My lattice results, however, indicate that $\widetilde{Z}_1$ 
indeed approaches a constant for large values of $\beta $. 
A detailed analysis of these issues is work in progress~\cite{blo02b}.  

A recent comprehensive study of the gluon propagator can be found 
in~\cite{bon01}. 
My numerical results for the pure SU(2) theory are shown in figure 
\ref{fig:1} and \ref{fig:2}. At high energies 
the lattice data for the form factors nicely reproduce the predictions 
from perturbation theory and, in particular, the correct anomalous 
dimensions. Most important, a clear signal of a diverging ghost 
propagator for the momentum $p \rightarrow 0$ is obtained 
(see figure \ref{fig:2}). 

Figure \ref{fig:1} also shows the running coupling constant. 
The red solid line is the one loop perturbative prediction which 
diverges at $\Lambda _{QCD}$, the Landau pole position. Comparing this 
1-loop fit with the lattice data, it is possible to estimate 
the scale $\Lambda _{QCD}$ of the one loop level which turns out to 
be roughly $1\, $GeV. 

\subsection{ The modified non-confining theory } 

\begin{figure}[tbh] 
\centerline{ 
\epsfxsize=6cm\epsffile{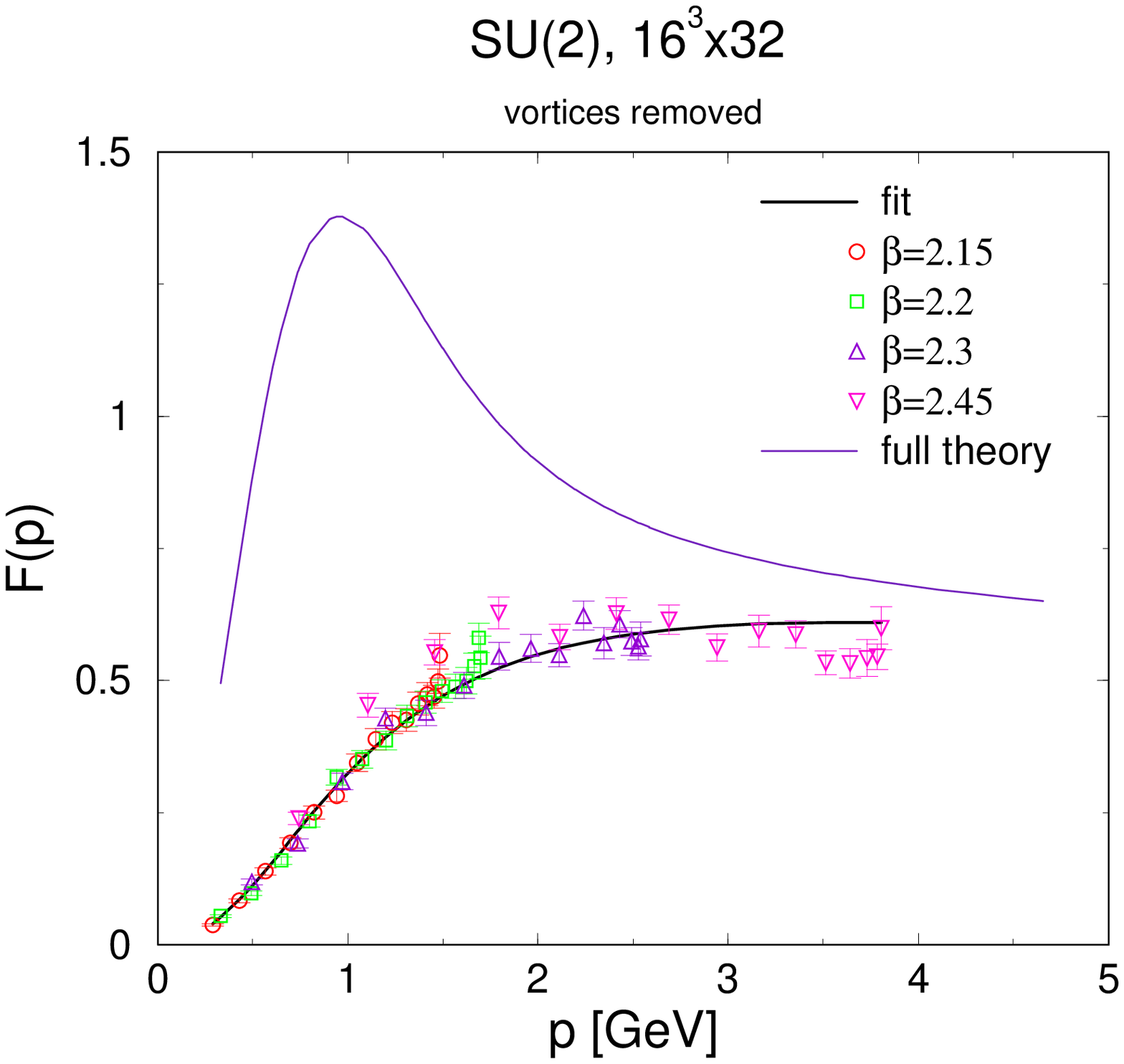}
\epsfxsize=5.7cm\epsffile{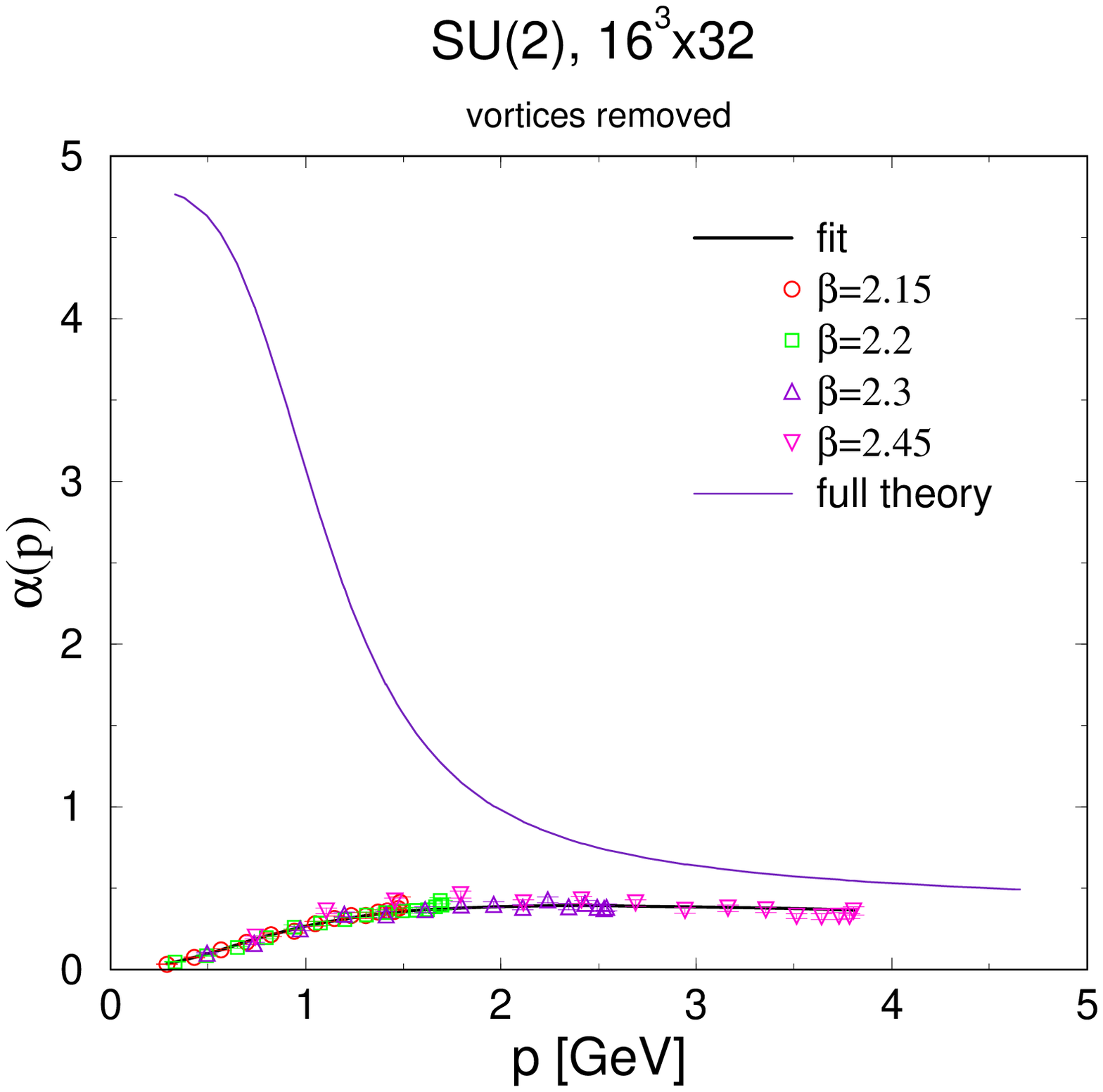}
}
\caption{ The gluon form factor (left panel) and the running coupling 
   strength (right panel) as function of the momentum transfer; modified 
   non-confining SU(2) theory.
}\label{fig:3}
\end{figure}
In order to construct the modified non-confining theory, I firstly 
implemented the maximal center gauge with the help of the 
procedure outlined in~\cite{deb97}. The $Z_\mu (x)$ center fields 
are constructed from the full link configuration $U_\mu (x)$ by 
center projection. The link variables $U^\prime _\mu (x) $ of the 
modified theory are subsequently defined by $
U^\prime _\mu (x) \; = \; Z_\mu (x) \; U_\mu (x)$. When 
calculating the static quark anti-quark potential using the modified 
configurations, one finds a Coulomb type of behavior. 

\begin{figure}[tbh] 
\centerline{ 
\epsfxsize=6.5cm\epsffile{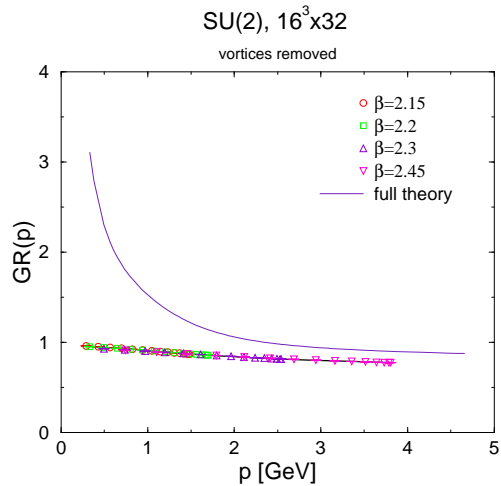}
}
\caption{ The ghost (right panel) form factor 
   as function of the momentum transfer; modified non-confining SU(2) theory.
}\label{fig:4}
\end{figure}
\vskip 0.3cm 
The modified configurations are subsequently subjected to the Landau 
gauge fixing procedure. Employing the Landau gauge fixed ensembles, 
one obtains the gluon and ghost form factor and the running 
coupling of the modified theory from the usual correlation functions. 
This procedure was carried out at several $\beta $ values. 
The result is shown in figures \ref{fig:3} and \ref{fig:4}. 
The solid lines are the fits to the data points obtained in pure 
SU(2) theory. One observes that the interaction strength in the 
intermediate momentum range is drastically reduced when 
the vortices have been removed. In addition, one observes that 
divergence of the ghost form factor in the infrared limit disappears. 
This indicates that in the Zwanziger picture of confinement the removal 
of the vortices shifts the relevant configurations away from the Gribov 
horizon: ghost and gluon form factors only acquire moderate corrections 
to the free field limit. 
Since the SU(2) scaling behavior also applies for the case of the 
modified theory, one can use (\ref{eq:3}) to remove the lattice 
spacing in favor of a physical energy scale: indeed, one observes that 
data points for the gluon and ghost form factor fall on top of a single 
curve also in the case of the modified theory. 

\vspace{.3cm}
{\bf Acknowledgments.} It is a pleasure to thank my collaborators 
\break 
J.~C.~R.~Bloch, A.~Cucchieri, J.~Gattnar, T.~Mendes, H.~Reinhardt. 
I thank the organizers for the invitation to this stimulating 
workshop.

\end{document}